\documentclass[prl,twocolumn,noshowpacs,preprintnumbers,amsmath,amssymb,nofootinbib,showkeys,superscriptaddress]{revtex4-2}
\usepackage{bbm}
\usepackage{lipsum}
\usepackage{mathrsfs}
\usepackage{graphicx}
\usepackage{dcolumn}
\usepackage{bm}
\usepackage{color}
\usepackage{comment}
\usepackage{hyperref}
\usepackage{times}
\usepackage{ulem}
\usepackage{footnote}

\newcommand{\tb}{\textbf}

\newcommand{\eps}{\epsilon}

\newcommand{\ii}{{\rm i}}

\newcommand{\dd}{{\rm d}}
\newcommand{\rmE}{{\mathrm{E}}}

\begin{document}
\title{
		Why financial prices exhibit Brownian motion despite predictable order
    flow: \\An exactly solvable nonlinear price impact model
}
\author{Yuki Sato}
\affiliation{Institute of Systems and Information Engineering, University of Tsukuba, Tennodai, Tsukuba, Ibaraki 305-8573, Japan}
\email{sato.yuki.gb@u.tsukuba.ac.jp}

\author{Shunta Fujiwara}
\author{Kiyoshi Kanazawa}

\affiliation{Department of Physics, Graduate School of Science, Kyoto University, Kyoto 606-8502, Japan}
\date{\today}
\begin{abstract}
	In econophysics, there are two enigmatic empirical laws of financial market microstructure: (i)~the market-order flow has predictable persistence due to metaorder splitting by institutional investors, as formulated by the Lillo--Mike--Farmer model. However, this phenomenon seems paradoxical given the diffusive and unpredictable price dynamics. (ii)~The price impact $I(Q_{\rm tot})$ of a large metaorder $Q_{\rm tot}$ follows the square-root law, $I(Q_{\rm tot})\propto \sqrt{Q_{\rm tot}}$. In this Letter, we reveal why price dynamics follows Brownian motion despite predictable order flow by unifying these enigmas via an exactly solvable model. We generalize the Lillo--Mike--Farmer model to nonlinear price-impact dynamics, which is mapped to a solvable L\'evy-walk model. Our exact solution shows that the price dynamics remains diffusive under the square-root law, even under persistent order flow, reflecting the finite variance of metaorder impact; our model is thus a minimal representative of the universality class of nonlinear price-impact models with finite-variance metaorder impact. This work illustrates the crucial role of the square-root law in mitigating price movements by large metaorders, leading to Brownian price dynamics, consistent with the efficient market hypothesis over long timescales.
\end{abstract}


\maketitle

\begin{figure*}
	\includegraphics[width=160mm]{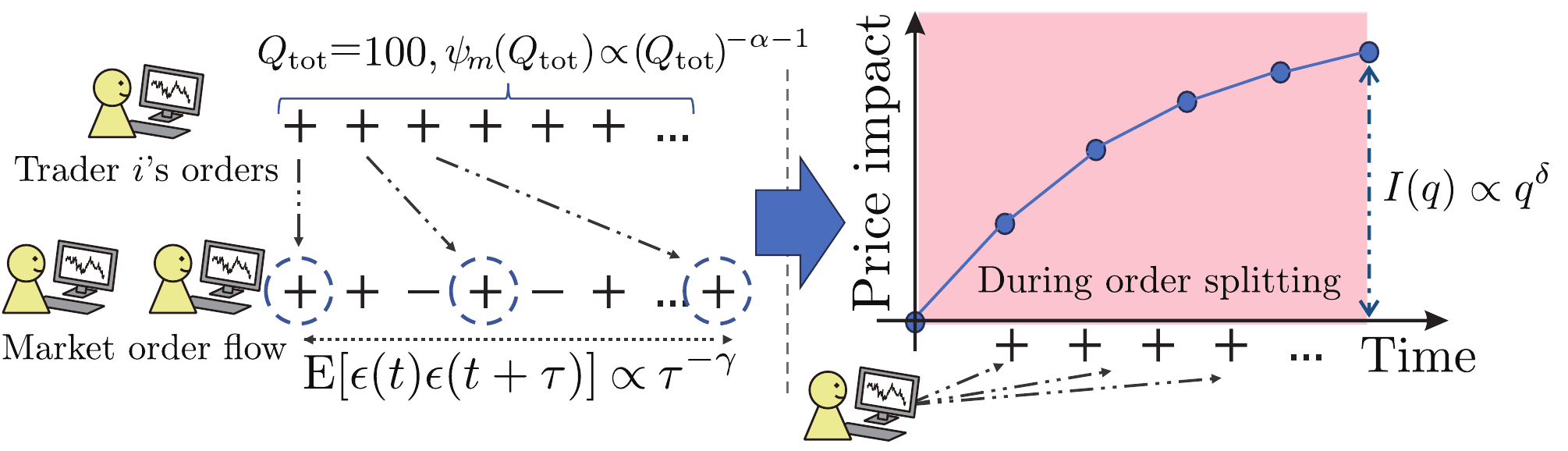}
	\caption{
		Order-splitting behavior underlying the Lillo--Mike--Farmer (LMF) model (left) and the corresponding price impact (right). Institutional traders typically split their large metaorders into a long sequence of small child orders to minimize their transaction cost. The long-term predictability (long memory) of market-order flow arises due to this order splitting because all child orders within the same metaorder have the same sign. The resulting market impact---the price change between the initial and final times of metaorders---empirically follows a nonlinear scaling $I(Q_{\rm tot})\propto \sqrt{Q_{\rm tot}}$, called the square-root law.
		}\label{fig:Schematic-OST-PI}
\end{figure*}

\paragraph*{Introduction.}
	There are two empirical enigmas in econophysics~\cite{BouchaudText,Cont2001,Slanina}. The first enigma is the long memory of the market-order flow. A market order is a decision to buy or sell stocks immediately, and its sign is represented by $\eps_t=+1$ ($\eps_t=-1$) for a buy (sell) market order at time $t$ (e.g., $\{\eps_t\}_{t\geq 1}=\{+,+,+,-,-,+\}$ signifies three consecutive buy orders, followed by two sell orders and then one buy order). The long memory implies that such order signs are easily predictable and, thus, the autocorrelation (ACF) of the market-order signs decays very slowly:
	\begin{equation}
		\rmE [\eps_t\eps_{t+\tau}]\propto \tau^{-\gamma}, \>\>\> 0<\gamma<1,
    \end{equation}
	where $\rmE[\dots]$ denotes an ensemble average over stationary realizations of the order-flow process and the sum of the ACF diverges as $\sum_{\tau =1}^\infty \rmE[\eps_t\eps_{t+\tau}]=\infty$. Empirically, the ACF is estimated by a time average under the assumptions of stationarity and ergodicity. This formula implies that the order flow has persistence for a very long time, such as for a few days~\cite{BouchaudText}.

	Then, why does this predictability arise ubiquitously in financial markets? This order-sign predictability originates from the metaorder splitters and is well formulated by the Lillo--Mike--Farmer (LMF) model~\cite{LMF_PRE2005,GeneralizedLMF}. The LMF theory argues that institutional investors have hidden large orders (called metaorders), which are split into a series of child orders. Therefore, large metaorders cause persistent order flow because all child orders have the same order sign during the execution of a single metaorder (see the left panel of Fig.~\ref{fig:Schematic-OST-PI}). In particular, the LMF theory predicted a relation between the market microstructure and the ACF,
	\begin{equation}
		\gamma=\alpha-1, \>\>\> 1<\alpha<2, 
	\end{equation}
	by assuming a power-law distribution $\psi_m(Q_{\rm tot})\propto Q^{-\alpha-1}_{\rm tot}$ for the total metaorder size $Q_{\rm tot}$. Recently, Sato and Kanazawa validated this prediction by scrutinizing the microscopic dataset of the Tokyo Stock Exchange (TSE)~\cite{SatoPRL2023,SatoPRR2023} and, thereby, have established the LMF theory as the causality theory for the long memory.
	
	However, this phenomenon is very counter-intuitive: given that the market-order flow is predictable, why is the price dynamics unpredictable? This puzzle arises because the price dynamics would exhibit predictable anomalous diffusion if the price impact were linear with respect to the metaorder size, as documented in the linear propagator model~\cite{BouchaudText,BouchaudProp1,BouchaudProp2} (except under the fine-tuned balance condition between the order-flow and price-impact decays~\cite{Maitrier2026}). Why is the price dynamics diffusive despite the predictable order flow? This is the first empirical enigma, known as the price diffusivity puzzle; it is particularly puzzling because most traditional economic theories predict linear price-impact models~\cite{Kyle1985}.

	The second enigma is the nonlinearity of the price impact, called the square-root law (SRL)~\cite{BouchaudText,Toth2011} (see Fig.~\ref{fig:Schematic-OST-PI}). In practice, the linear price-impact model is valid only for small metaorder size $Q_{\rm tot}$; for large $Q_{\rm tot}$, it has been reported that the price impact is a nonlinear function of the metaorder size $Q_{\rm tot}$: 
	\begin{equation}
		I(Q_{\rm tot}):=\rmE [\eps \Delta m \mid Q_{\rm tot}] \propto Q_{\rm tot}^{\delta}, \>\>\> \delta\approx \frac{1}{2},
	\end{equation}
	where $\Delta m$ is the price change caused by a metaorder of total size $Q_{\rm tot}$. Recently, Sato and Kanazawa scrutinized the TSE dataset, provided a very accurate estimation of $\delta$, and established the strict universality of the SRL, such that $\delta$ is exactly equal to $1/2$ for all liquid stocks on the TSE within statistical errors~\cite{SatoPRL2025}. To date, the cause of the strict universality of the SRL remains unclear. One of the most promising models might be the latent limit order-book model~\cite{BouchaudText,Toth2011,DonierLOB2015}, but partially negative evidence was observed recently~\cite{Guillaume2025}. Thus, there is no consensus yet regarding the microscopic origin of the SRL.

	\begin{figure*}
		\includegraphics[width=120mm]{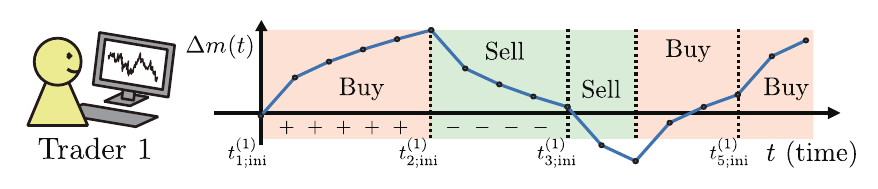}
		\caption{
			LMF L\'evy-walk model for the single-trader case $M=1$. Splitter~1 sequentially executes signed metaorders, where all child orders within each metaorder share the same sign. The total metaorder size $Q_{\rm tot}$ and the resulting price impact $I(Q_{\rm tot})$ follow the power-law distribution $\psi_m(Q_{\rm tot})\propto Q_{\rm tot}^{-\alpha-1}$ and the nonlinear scaling $I(Q_{\rm tot})\propto Q_{\rm tot}^{\delta}$, respectively. The square-root law corresponds to $\delta=1/2$. The cumulative price change $\Delta m(t):=m(t)-m(0)$ is the sum of the signed nonlinear price-impact contributions of successive metaorders. This process is a discrete-time L\'evy walk with nonlinear space--time coupling. For $M>1$, the total price change is the sum of the corresponding contributions from all splitters.
		}\label{fig:model}
	\end{figure*}
	In this Letter, together with the companion paper~\cite{SFK2026}, we propose a minimal model that naturally extends the LMF framework to incorporate nonlinear price-impact dynamics, thereby unifying these two enigmas from a single, coherent perspective. We consider the system composed of $M$ order splitters whose metaorder-size statistics follow the power law $\psi_m(Q_{\rm tot})\propto Q_{\rm tot}^{-\alpha-1}$ with $1<\alpha<2$. The price impact of each metaorder is assumed to obey the nonlinear scaling $I(Q_{\rm tot})\propto Q^{\delta}_{\rm tot}$ with $0<\delta\leq 1$. Crucially, this model is exactly solvable. The price-impact contribution by a single trader can be mapped to the L\'evy-walk theory~\cite{KlafterB,gLWPRL1987} with nonlinear space--time coupling~\cite{KlafterPRA1987}. Our exact solution shows that the price dynamics is always diffusive even in the presence of the long memory if $\delta=1/2$. This result implies that the SRL plays the crucial role in mitigating the price impact due to the huge liquidity consumption by institutional investors. Additionally, we find that natural extensions of our model exhibit behavior consistent with other empirical laws (such as the inverse-cubic law~\cite{BouchaudText,Slanina,GabaixNature,Gabaix2006,ICLLetter,ICLCompany,ICLIndex} and volatility clustering~\cite{Slanina,Cont2001}). Thus, our minimal model is consistent with various empirical laws in econophysics.
    
\paragraph{LMF Model.}
    We start from the Lillo--Mike--Farmer (LMF) model with $M$ statistically independent order splitters indexed by $\Omega=\{1,2,\ldots,M\}$. Each splitter repeatedly executes signed metaorders of total size $Q^{(i)}_{\rm tot}$, where $Q^{(i)}_{\rm tot}$ is randomly drawn from the empirically observed power-law total metaorder-size distribution $\psi_m(Q_{\rm tot})$:
    \begin{equation}
        \psi_{m}(Q_{\rm tot})\propto (Q_{\rm tot})^{-\alpha-1},\quad1<\alpha<2.
    \end{equation}
    At each time step, one splitter $i\in\Omega$ is randomly selected with equal probability $\lambda:=1/M$ and submits a unit child order. During a metaorder execution, all child orders share the same order sign $\eps^{(i)}$. Once the current metaorder is completed, the splitter independently draws a new sign $\eps^{(i)}\in\{-1,+1\}$ with equal probability and a new total size $Q_{\rm tot}^{(i)}$. In this Letter, we extend the original LMF framework by incorporating a price process driven by nonlinear price impact.

\paragraph{Price dynamics for $M=1$.}
    Let us introduce a price process into the LMF model. To develop our intuition, we first describe the single-splitter dynamics and then extend it to the full $M$-body system. 
		
	For $M=1$, let $q^{(1)}(t)$ be the volume already executed immediately before time $t$ within the current metaorder of splitter~1. This time-dependent variable should be distinguished from the total metaorder size $Q^{(1)}_{\rm tot}$. The counter $q^{(1)}(t)$ is incremented by one unit at each execution. Once it reaches the metaorder size $Q^{(1)}_{\rm tot}$, the current metaorder is completed; then $q^{(1)}(t)$ is reset to zero, and a new sign and a new total size $Q^{(1)}_{\rm tot}$ are drawn for the next metaorder. In this case, splitter~1 deterministically submits a child order at every time step.

    We assume that the price is driven by the price impact generated by metaorder splitting (see Fig.~\ref{fig:model} for a schematic). Namely, when splitter $1$ submits a child order at time $t$, the price $m(t)$ is incremented as
    \begin{equation}
		\begin{gathered}
            m(t+1)-m(t)= \eps^{(1)}(t)\Delta I\left(q^{(1)}(t)\right),\\
            \Delta I(q):=I(q+1)-I(q),
		\end{gathered}
        \label{eq:price_dynamics}
    \end{equation}
    where $I(q)$ denotes the price impact generated by the first $q$ child orders within the same metaorder. Thus, $\Delta I(q)$ is the price-impact increment caused by the child order. We assume that this path-impact function follows the empirically observed power-law form,
    \begin{equation}
        I(q)=c_0 q^\delta, \quad 0<\delta\leq 1,
        \label{eq:nonlinear_impact}
    \end{equation}
    where $c_0>0$ sets the price scale. Empirical studies support the validity of the SRL scaling even along execution paths~\cite{BouchaudText}. At completion, $q=Q_{\rm tot}$, and hence the total impact is $I(Q_{\rm tot})$. In the following, we set $c_0=1$ without loss of generality by choosing the price unit appropriately. The case $\delta=1$ corresponds to linear impact, whereas the empirically relevant square-root impact law corresponds to $\delta=1/2$.

    Equivalently, accumulating the price-impact increments over successive metaorders gives
	\begin{subequations}
		\label{eq:price-dynamics}
    	\begin{align}
    		\Delta m(t) &:=m(t)-m(0) 
            = \sum_{k=1}^{N^{(1)}(t)} \eps_{k}^{(1)}\left(q_k^{(1)}(t)\right)^{\delta},\\
            q_k^{(1)}(t)&:=\min\{ t_{k;{\rm fin}}^{(1)},t\}-t_{k;{\rm ini}}^{(1)}.
    	\end{align}
	\end{subequations}
    Here, $t_{k;{\rm ini}}^{(1)}$, $t_{k;{\rm fin}}^{(1)}$, and $\eps_k^{(1)}$ denote the initial time, final time, and order sign of the $k$-th metaorder of splitter $1$, respectively, and $N^{(1)}(t)$ is the number of metaorders initiated by time $t$. The quantity $q_k^{(1)}(t)$ represents the executed volume of the $k$-th metaorder accumulated up to time $t$.

\paragraph{L\'evy-walk mapping}
		This model~\eqref{eq:price-dynamics} can be regarded as a discrete-time L\'evy-walk model with nonlinear space-time coupling. Indeed, the price impact of each metaorder corresponds to one flight: the elapsed execution time corresponds to the flight time, the order sign determines the flight direction, and the signed price displacement generated during the flight is given by $\pm(\Delta t)^\delta$. Thus, the nonlinear price-impact function realizes a nonlinear coupling between flight time and displacement. We therefore refer to the model~\eqref{eq:price-dynamics} as the {\it discrete-time LMF L\'evy-walk model}.
		
		This discrete-time LMF L\'evy-walk model can be readily extended to a continuous-time counterpart. Both models are exactly solvable within the L\'evy-walk framework and exhibit the same asymptotic scaling of the mean-squared displacement (MSD) and the same correlation exponent. Therefore, the exact solution of the continuous-time model is outlined in the End Matter, while detailed derivations for both the discrete- and continuous-time models are provided in our companion paper~\cite{SFK2026}.

\paragraph{Price dynamics for general $M$.}
    Let us consider the $M$-body dynamics as a natural generalization of the above single-trader dynamics. The corresponding price dynamics for the $M$-splitter system is obtained by accumulating the price-impact contributions from all splitters:
    \begin{equation}
            \Delta m(t) = \sum_{i\in\Omega} \Delta m^{(i)}(t) 
            = \sum_{i\in\Omega}\sum_{k=1}^{N^{(i)}(t)} \eps_{k}^{(i)}\left(q_{k}^{(i)}(t)\right)^{\delta},
		\label{eq:multi-splitter-price-dynamics}
    \end{equation}
    where $q_k^{(i)}(t)$ denotes the cumulative volume of the $k$-th metaorder executed by splitter $i$ up to time $t$, and $N^{(i)}(t)$ is the number of metaorders initiated by splitter $i$ up to time $t$. The variable $q_k^{(i)}(t)$ increases only when splitter $i$ is selected. The quantity $\Delta m^{(i)}(t)$ denotes the price-impact contribution of splitter $i$ to the total price displacement. 

	In this work, we employ deterministic price impact~\eqref{eq:nonlinear_impact} as a minimal assumption. Note that stochastic fluctuations in individual price-impact increments or other empirical ingredients (such as resting periods between metaorders and post-metaorder impact decay) can also be introduced as further generalizations. However, these extensions do not affect the central results of this Letter, as shown in our companion paper~\cite{SFK2026}. We therefore consider this simplest formulation as a minimal resolution of the price diffusivity puzzle addressed here.

\paragraph{Brownian price dynamics under the square-root law.}
	\begin{figure*}
		\includegraphics[width=180mm]{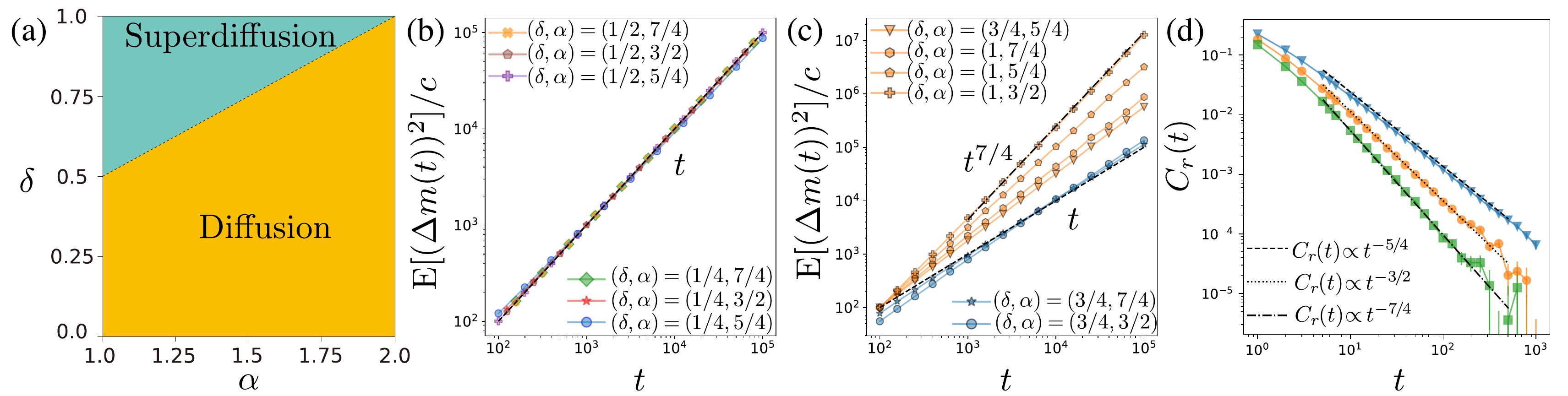}
		\caption{			
			(a)~Phase diagram between superdiffusion and normal diffusion. The phase boundary is given by $2\delta=\alpha$, implying that the price dynamics is always diffusive under the SRL $\delta=1/2$.
			(b)~Numerical mean-squared displacement (MSD) for $\delta\in \{0.25,0.5\}$, $\alpha\in \{1.25,1.5,1.75\}$, and $M=1$, showing normal diffusion. The constant $c$ denotes the prefactor estimated from the numerical MSD.
			(c)~Numerical MSD for $\delta\in \{0.75,1\}$, $\alpha\in\{1.25,1.5,1.75\}$, and $M=1$, illustrating the transition from normal diffusion to superdiffusion across the phase boundary $2\delta=\alpha$.
            (d)~Numerical return ACF for $\delta=0.5$, $\alpha\in\{1.25,1.5,1.75\}$, and $M=1$. The numerical results are consistent with the predicted decay $C_{r}(\tau)\propto \tau^{-(\alpha-2\delta+1)}$.
		}\label{fig:diffusion}
	\end{figure*}
	Let us present our first main result. From the exact solution, we obtain the following asymptotic scaling for the MSD:
	\begin{equation}
		\rmE[ \Delta m^2(t) ] \propto 
		\begin{cases}
			t^{1+2\delta-\alpha} & \mbox{if $2\delta > \alpha$} \\
			t & \mbox{if $2\delta < \alpha$}
		\end{cases},
		\label{eq:MSD}
	\end{equation}
	implying that superdiffusion arises when $2\delta > \alpha$. $\rmE[\cdots]$ denotes an ensemble average over realizations of the stochastic price process, including metaorder sizes, order signs, and splitter selections. Under the standard LMF assumption $1<\alpha <2$, the price dynamics always exhibits normal diffusion for $\delta \leq 1/2$ (see Fig.~\ref{fig:diffusion}). See End Matter for the derivation. At the boundary $2\delta=\alpha$, a logarithmic correction appears. Note that this result is consistent with the linear propagator model~\cite{BouchaudProp1,BouchaudProp2} for the specific case with non-zero permanent impact for $\delta=1$.

	This result can be intuitively understood from the central limit theorem. Consider the finite-variance condition for the metaorder-level permanent impact: 
	\begin{equation}
		\rmE[I^2(Q_{\rm tot})] \propto \int_1^\infty Q_{\rm tot}^{-\alpha+2\delta-1} \dd Q_{\rm tot} < \infty.
	\end{equation}
	This condition is satisfied if and only if $2\delta < \alpha$, which naturally leads to diffusive price dynamics due to the central limit theorem. The exact solution~\eqref{eq:MSD} is consistent with this simple mechanism. Our argument should essentially apply to any nonlinear price-impact model provided that independent metaorders have finite non-zero variance of permanent impact. Our L\'evy-walk model is the minimal representative of the universality class of nonlinear price-impact models with finite-variance metaorder impact.

\paragraph{Short-memory price dynamics under the square-root law.}
	Furthermore, the SRL suppresses long-range predictability in price dynamics. Let us consider the continuous-time velocity $v(t):=\dd m(t)/\dd t$ as the counterpart of the discrete-time return process. In the stationary state, the velocity correlation asymptotically behaves as 
	\begin{equation}
		C_{r}(\tau):=\rmE[v(t)v(t+\tau)]_{\rm ss}\propto\tau^{-\theta},\>\>\>\theta=\alpha-2\delta+1.
	\end{equation}
	When the price impact obeys the SRL, $\delta=1/2$, the decay exponent becomes $\theta=\alpha>1$. The price-change correlation is short-ranged, in sharp contrast to the long-ranged order-sign ACF with exponent $\gamma=\alpha-1<1$. Thus, the SRL suppresses not only anomalous diffusion but also long-range memory in the price dynamics.
	
	These two results suggest that the market exhibits strong resilience against large liquidity consumption by order splitters thanks to the sufficiently concave nature of the SRL. This interpretation is crucially important: it is the SRL that suppresses the large price movements in the presence of the long memory. This scenario highlights the importance of studying the microscopic origin of the SRL for stable regulation of financial markets. 

\paragraph{Concluding discussion.}
	We develop an exactly solvable model of nonlinear price-impact dynamics in the presence of metaorder splitters, as a minimal representative of the universality class with finite-variance metaorder impact. Our theory clarifies the practical importance of the SRL regarding market stability. The presence of the long memory implies that markets suffer from the large liquidity consumption by order splitters. It has been mysterious why markets operate consistently with the diffusive nature even in the presence of such a large demand. Our theory shows that the concavity of the SRL suppresses large price movements caused by order splitters, which leads to diffusive price dynamics with short memory. In other words, if the price impact were less concave than the SRL ($\delta>1/2$), the price would be superdiffusive with clear predictability for $1<\alpha<\alpha^*:=2\delta$.
	
	Our theory highlights that the SRL is not a minor technical topic but a key topic in financial economics, directly related to the theoretical foundation of the efficient market hypothesis (EMH)---the hypothesis that the price is unpredictable (thus diffusive) from public information. However, most traditional economic theories (such as Kyle's model~\cite{Kyle1985}) predict the linear price-impact law, which apparently contradicts the EMH in the presence of the predictable market-order flow. Our theory clarifies that, even under predictable market-order flow, the long-timescale price dynamics is consistent with the EMH thanks to the SRL. There is room to update financial-economic theories by incorporating the SRL.

	Our model also serves as a minimal null model questioning the previous statistical evidence for fine-tuned impact decay. Within the linear propagator model, the fine-tuned impact decay $G(\tau)\propto\tau^{-\beta_c}$, $\beta_c:=(1-\gamma)/2$, has been proposed as a resolution of the diffusivity puzzle~\cite{BouchaudProp1,BouchaudProp2}, and propagators inferred from data have been reported to be consistent with it~\cite{BouchaudText}. However, we find that the same propagator estimation applied to our model returns the fine-tuned decay as a statistical artifact~\cite{FSK}, although our model has no true impact decay. More generally, for diffusive prices and long-memory order flow, propagator estimation returns the fine-tuned decay irrespective of the true impact~\cite{FSK}, so that the inferred decay alone is not evidence for transient impact. Thus, our model provides a minimal counterexample to the causal reading of the inferred fine-tuned decay as true impact decay.

	As documented in our companion paper~\cite{SFK2026}, the present framework can naturally accommodate additional empirical ingredients, such as resting periods between metaorder executions and post-metaorder impact decay. These extensions do not change the central conclusions concerning diffusion and short memory of this Letter. Furthermore, such extended formulations may help explain broader empirical regularities of financial prices. For example, models incorporating resting periods or impact decay exhibit behavior consistent with the inverse-cubic law of price changes~\cite{BouchaudText,Slanina,GabaixNature,Gabaix2006,ICLLetter,ICLCompany,ICLIndex} and volatility clustering~\cite{Slanina,Cont2001}. We mention these observations here as further evidence that metaorder splitting combined with nonlinear price impact provides a useful mesoscopic framework for connecting several empirical regularities in financial markets. For a more quantitative understanding of financial markets, it would be natural to incorporate market-microstructure dynamics and time-dependent endogenous/exogenous parameters, such as self-excitation mechanisms~\cite{BlancQF2017,KzDiderPRL2021,KzDiderPRR2023}, limit-order-book dynamics, intraday seasonality, and external shocks. Such extensions are left as future studies.

	\begin{acknowledgments}
		We thank J.-P. Bouchaud and G. Maitrier for fruitful comments. YS was supported by JSPS KAKENHI (Grant No. 24KJ1328 and 26K24638). KK was supported by JSPS KAKENHI (Grant Nos. 21H01560, 22H01141, and 25K01450 and the JSPS Core-to-Core Program (Grant No. JPJSCCA20200001).
	\end{acknowledgments}

	\section*{Author contribution}
		YS developed the numerical program code and contributed to the analytical calculations for the mean-squared displacement. SF contributed to the theoretical analysis of the discrete model. KK designed and managed the project, and contributed to the analytical calculations. All authors contributed to writing the manuscript and approved its final content.

	\section*{Code availability}
		Simulation codes are provided in the Supplemental Material.

\clearpage

\section{End Matter}

	The outline of our theoretical calculations is given here. We simplify the notation based on Ref.~\cite{KlafterB} by rewriting the contribution of trader $i$, $\Delta m^{(i)}(t)$, as the position of a L\'evy-walk particle $X(t)$. We use $x$ for a single-flight displacement and $X$ for the cumulative position. More detailed calculations are presented in the companion paper~\cite{SFK2026}. 

\subsection{Appendix A: The discrete-time single-splitter model}
	For the case $M=1$, we formulate the discrete-time LMF price-impact dynamics~\eqref{eq:price-dynamics} within the discrete-time L\'evy-walk framework. Since the size and sign of each metaorder are independently renewed after the completion of the previous metaorder, each metaorder naturally corresponds to one flight of a L\'evy walk. In this mapping, the metaorder execution time corresponds to the flight time, the order sign determines the flight direction, and the total price displacement generated during one metaorder of duration $t$ is given by $x=\pm t^\delta$. Accordingly, the corresponding space-time coupling function is written as
		\begin{equation}
				\psi(x,t)=\frac{1}{2}\delta(|x|-t^{\delta})\psi_{m}(t), \quad
				\psi_m(t)=\frac{t^{-\alpha-1}}{\zeta(\alpha+1)},
		\end{equation}
	where the flight time $t$ is a positive integer, $1<\alpha<2$, and $\zeta(s):=\sum_{n=1}^{\infty}n^{-s}$ is the Riemann zeta function. The power-law metaorder-size distribution underlying $\psi_m(t)$ and the square-root impact law corresponding to $\delta=1/2$ are empirically supported by Refs.~\cite{SatoPRL2023} and \cite{BouchaudText,SatoPRL2025}, respectively. The cumulative displacement over successive flights is denoted by $X(t)$, corresponding to the price displacement $\Delta m(t)$ in our LMF price-dynamics model. This mapping allows us to apply the standard renewal formulation of L\'evy-walk theory to the single-splitter price process.

	This formulation can be extended to the multi-trader case $M>1$, where one trader is selected with probability $1/M$ at each time step, executes a child order, and causes the corresponding price-impact increment $\Delta I$. The exact formulation and solution are presented in the companion paper~\cite{SFK2026}. 
	
	In this End Matter, we outline our theoretical calculations only for the continuous-time model, since the discrete-time LMF L\'evy-walk model and its continuous-time counterpart exhibit the same asymptotic MSD scaling and short-memory property.

	\begin{figure*}[t]
		\includegraphics[width=170mm]{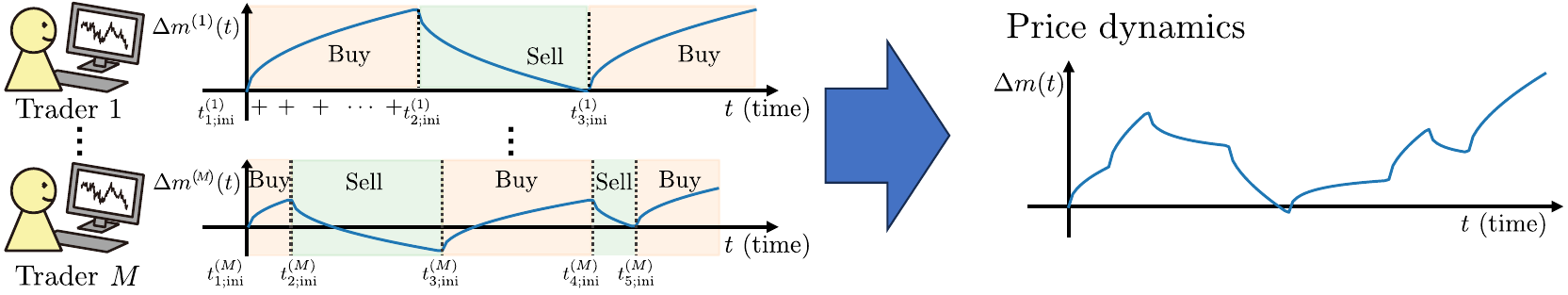}
		\caption{
			Extension of the single-trader model ($M=1$) to the multi-trader model ($M>1$). All traders make their decisions independently, and their metaorders cause the corresponding price-impact contributions $\Delta m^{(i)}(t)$. The total price displacement is their sum, $\Delta m(t)=\sum_{i=1}^M\Delta m^{(i)}(t)$.
			}
		\label{fig:modeldynamics-mbody}
	\end{figure*}

\subsection{Appendix B: Continuous-time L\'evy walk}
	    We now consider the continuous-time counterpart of the discrete-time single-splitter model. In the discrete model, the executed volume of the current metaorder increases by one unit whenever the splitter submits a child order:
    \begin{equation}
	        q(t)\mapsto q(t)+1,\quad 0\leq q(t)<Q_{\rm tot}.
    \end{equation}
    In the continuous-time representation, this unit-rate execution is replaced by continuous evolution at a constant rate:
    \begin{equation}
	   \frac{\dd q(t)}{\dd t}=\nu,\quad\nu=1, \quad 0\leq q(t)<Q_{\rm tot}.
    \end{equation}
	    When $q(t)$ reaches the total metaorder size $Q_{\rm tot}$, the metaorder is completed; then $q(t)$ is reset to zero, and a new total size $Q_{\rm tot}$ and a new sign $\eps=\pm1$ are drawn independently. Thus, during a metaorder that starts at $t_{\rm ini}$, we have
    \begin{equation}
		   q(t)=t-t_{\rm ini}, \quad t_{\rm ini}\leq t<t_{\rm fin},
    \end{equation}
	    where $t_{\rm fin}-t_{\rm ini}=Q_{\rm tot}$. Hence, for a metaorder with duration $\Delta t$, the total executed volume is $Q_{\rm tot}=\Delta t$, and the price displacement generated by this metaorder is $x=\eps(\Delta t)^\delta$.

	    This continuous-time model keeps the metaorder sign-generation mechanism of the discrete LMF model: within each metaorder, the sign $\eps$ remains fixed, whereas, after completion, a new duration and a new sign are drawn independently. Therefore, each metaorder execution can be identified with a single L\'evy-walk flight: the duration $\Delta t$ is the flight time, the sign $\eps$ determines the flight direction, and $\eps(\Delta t)^\delta$ is the signed flight displacement. Accordingly, the joint density of the flight displacement $x$ and flight time $t$ is
    \begin{equation}
	        \psi(x,t) = \frac{1}{2} \delta\!\left(|x|-t^\delta\right) \psi_m(t),\quad \psi_m(t)=\alpha t^{-\alpha-1}\Theta(t-1),
	    \end{equation}
	  where $\Theta(t)$ is the Heaviside step function. This joint density represents the nonlinear space--time coupling generated by the nonlinear price-impact law.

	Let $\eta(X,t)$ be the probability density that the walker completes a flight exactly at time $t$ and arrives at position $X$. This renewal density satisfies
	\begin{align}
		\eta(X,t)=\delta(X)\delta(t)+\int_0^t\dd t'\int\dd y\,\eta(y,t')\psi(X-y,t-t').
	\end{align}
	Using $\eta(X,t)$, the probability density function (PDF) $P(X,t)$ that the walker is at position $X$ at time $t$ is written as
	\begin{align}
		P(X,t) = \int \dd y \int_{0}^{t}\dd t' \eta(y,t')\Psi(X-y,t-t'),
	\end{align}
	where $\Psi_m(t):=\int_{t}^{\infty}\dd t'\psi_m(t')$ and $\Psi(x,t):=(1/2)\delta(|x|-t^{\delta})\Psi_m(t)$. Using the Fourier--Laplace representation
	\begin{equation}
		f(k,s):=\int_0^{\infty}\dd t\,e^{-st}\int_{-\infty}^{\infty}\dd X\,e^{-\ii kX}f(X,t),
	\end{equation}
	we obtain
	\begin{align}
		P(k,s) &= \frac{\Psi(k,s)}{1-\psi(k,s)}, \label{eq:app:cont_exact_pdf}\\
		\psi(k,s) &=  \int^{\infty}_{0}\dd t\,e^{-st}\psi_m(t)\cos(kt^\delta),\\
		\Psi(k,s) &=  \int^{\infty}_{0}\dd t\,e^{-st}\Psi_m(t)\cos(kt^\delta).
	\end{align}
	This gives the exact Fourier--Laplace representation of the single-splitter PDF.

    Let us expand $P(k,s)$ for small $k$ and $s$. We obtain 
	\begin{align}
		P(k,s) \approx s^{-1} -\frac{k^2}{2}\left(A_1s^{\alpha-2\delta-2}+B_1s^{-2} \right)
	\end{align}
	with some coefficients $A_1$ and $B_1$. The MSD is obtained from
	\begin{equation}
			\rmE[X^2(t)]=\mathcal{L}^{-1}_{s\to t}\left[-\left.\frac{\partial^2P(k,s)}{\partial k^2}\right|_{k=0}\right],
	\end{equation}
	where $\mathcal{L}^{-1}_{s\to t}$ denotes the inverse Laplace transform from the Laplace variable $s$ to time $t$. This gives the asymptotic MSD behavior~\eqref{eq:MSD}.

\subsection{Appendix C: Extension to the $M$-splitter system}
	The extension from the single-splitter model to the full $M$-splitter model is straightforward for the continuous-time model by assuming that the splitter contributions are statistically independent. Let $\Delta m^{(i)}(t)$ be the price-impact contribution generated by splitter $i$ (see Fig.~\ref{fig:modeldynamics-mbody}). The total price displacement is defined by
	\begin{equation}
		\Delta m(t)=\sum_{i=1}^{M}\Delta m^{(i)}(t).
	\end{equation}
	Therefore, the $M$-splitter PDF is the $M$-fold convolution of the corresponding single-splitter PDF. Equivalently, its characteristic function satisfies:
	\begin{equation}
		P_M(k,t) = \prod_{i=1}^{M}P_1^{(i)}(k,t)=\left[P_1(k,t)\right]^M.\label{eq:cf:m-body}
	\end{equation}
  Using the characteristic function of the price in the $M$-body system~\eqref{eq:cf:m-body}, we obtain
	\begin{equation}
			\rmE[\Delta m^2(t)] = \sum_{i=1}^{M}\rmE[(\Delta m^{(i)}(t))^2] = M \rmE[(\Delta m^{(1)}(t))^2].
	\end{equation}
	Thus, the number of splitters changes only the prefactor of the MSD, not its scaling exponent. Finally, the diffusion condition derived from the single-splitter problem remains valid even for the $M$-splitter system. See also our companion paper~\cite{SFK2026} for the discrete-time formulation.

\subsection{Appendix D: Velocity correlation in the continuous-time model}
    Let us study the velocity correlation of the price dynamics. In the continuous-time representation, the velocity $v(t):=\dd m(t)/\dd t$ is the counterpart of the one-step return in the discrete-time model. For the single-splitter process, the velocity correlation follows from the renewal velocity process studied in Ref.~\cite{MeyerBarkaiKantz2017}. In the stationary state, it asymptotically behaves as
    \begin{align}
	        \rmE\!\left[v^{(1)}(t)v^{(1)}(t+\tau)\right]_{\rm ss}
        \propto \tau^{-\theta},
        \quad
        \theta = \alpha - 2\delta + 1 .
        \label{eq:velocity-ACF-single}
    \end{align}
    
    The extension to the $M$-splitter system is straightforward because different splitter contributions are statistically independent for the continuous-time model. Since the total velocity is $v_M(t)=\sum_{i=1}^{M}v^{(i)}(t)$, we obtain
    \begin{align}
	        \rmE\!\left[v_M(t)v_M(t+\tau)\right]_{\rm ss}
        &=
        \sum_{i,j=1}^{M}
	        \rmE\!\left[v^{(i)}(t)v^{(j)}(t+\tau)\right]_{\rm ss} \label{eq:velocity-ACF-M}\\
        &=
        \sum_{i=1}^{M}
	        \rmE\!\left[v^{(i)}(t)v^{(i)}(t+\tau)\right]_{\rm ss}
        \propto \tau^{-\theta},\notag        
    \end{align}
		with $\theta = \alpha - 2\delta + 1$. Here, the cross terms with $i\neq j$ vanish because different splitters are independent and the order signs have zero mean. Under the SRL, $\delta=1/2$, the exponent becomes $\theta=\alpha>1$. Therefore, the velocity correlation is short-ranged, in contrast to the long-ranged order-sign ACF with exponent $\gamma=\alpha-1<1$.

\end{document}